PHYSICS

# Mass-invariant universal optical conductivity from quantum geometry

**Chang-geun Oh[1]*†, Sun-Woo Kim[2,3]*†, Kun Woo Kim[4]*, Bartomeu Monserrat[2]*, Jun-Won Rhim[5]***

**Mass is a defining property of particles, shaping their fundamental nature and interactions. In condensed matter systems, the effective mass of electrons has long been regarded as a key factor influencing material properties, including their transport and optical responses. In this work, we challenge this conventional wisdom by unveiling a mass-invariant universal optical conductivity, purely derived from quantum geometry, in quadratic band–touching semimetals. Specifically, the optical conductivity simplifies to $\sigma = \left(e^2/8\hbar\right)d_{max}^2$, independent of effective mass and other band structure details, where $d_{max}$ represents the maximum Hilbert-Schmidt quantum distance. Furthermore, under time-reversal and rotational symmetries, $d_{max}$ is restricted to discrete values of 0 or 1, leading to a quantized universal optical conductivity. We also use first-principles calculations to demonstrate the mass-invariant universal optical conductivity across multiple materials, including bilayer graphene, monolayer bismuth, monolayer kagome palladium thiophosphate, and other realistic material candidates. Our work establishes a previously unidentified class of universal quantities in quantum materials entirely governed by quantum geometry.**

## INTRODUCTION

Quantum geometry stands at the forefront of condensed matter physics, deepening our understanding of diverse quantum phenomena (*1*–*18*). The geometry of quantum states in a parameterized Hilbert space is described by the Hilbert-Schmidt quantum distance (*19*, *20*), which converts the resemblance between two quantum states into a positive number between 0 and 1. This naturally leads to the concept of the quantum geometric tensor, with its symmetric real part defining the quantum metric and its antisymmetric imaginary part corresponding to the Berry curvature (*19*–*22*). While the Berry curvature has been extensively explored because of its role in determining the topological properties of materials (*23*, *24*), only recently have studies highlighted the notable impact of the quantum metric on material properties. For instance, a nontrivial quantum metric is associated with the superfluid weight in superconductors (*1*, *3*, *25*), anomalous Landau level spreading (*26*, *27*), exciton size (*28*), superfluid weight of exciton condensate (*9*, *29*), polarizability (*30*–*32*), the quantum Hall effect in bilayer graphene (*33*), scattering processes of electrons in the presence of disorder (*34*), and bulk-interface correspondence in singular flat band systems (*35*, *36*).

The influence of quantum geometry on the optical properties of solids has gained substantial attention in recent years, revealing a fundamental connection between optical responses and quantum geometry (*37*–*43*). Massive Dirac fermion systems serve as prime examples, illustrating optical phenomena governed by quantum geometry. In these systems, the mass plays a crucial role in determining both the topological and geometric properties, as the Berry curvature and quantum metric explicitly depend on it. The mass endows the system with finite Berry curvature, which in turn determines the associated optical properties (*44*, *45*). Moreover, it modifies the quantum metric, thereby influencing the optical response, including both linear and higher-order nonlinear optical conductivities (*37*, *41*–*43*). Specifically, the sign and magnitude of the mass can result in significant changes in the intensity of linear optical conductivity (*41*, *42*) while also causing a sign change in the third-order photovoltaic Hall conductivity (*43*). For these reasons, the mass has been regarded as a critical tuning parameter that encodes material-specific details and controls both topological and geometric properties. However, this dependence complicates isolating geometric quantities from optical measurements, as the measured signal inherently mixes quantum geometry with material-specific parameters such as mass. Therefore, finding a universal transport or optical quantity, determined solely by quantum geometric properties and independent of material-specific band structure details, would provide a powerful tool for directly probing the geometric structure of Bloch wave functions.

In this work, we demonstrate universal linear optical conductivity in two-dimensional (2D) systems exhibiting well-isolated quadratic band touching (QBT). The optical conductivity depends solely on the geometric quantity $d_{max}$, independent of material-specific details, including effective masses and band velocities. Here, $d_{max}$ is the maximum value of the Hilbert-Schmidt distance between all possible pairs of Bloch wave functions within the same band around the touching point. Furthermore, we show that time-reversal and $n$-fold rotational symmetries enforce $d_{max} = 0$ or 1, leading to a quantized universal optical conductivity. Last, we perform first-principles calculations on various realistic materials with QBTs, confirming both the quantized universal optical conductivity of $e^2/(8\hbar)$ and its robustness under perturbations. Our work establishes a universal quantity purely derived from quantum geometry, providing a compelling way to directly probe quantum geometry.

[1]Department of Applied Physics, The University of Tokyo, Tokyo 113-8656, Japan. [2]Department of Materials Science and Metallurgy, University of Cambridge, 27 Charles Babbage Road, Cambridge CB3 0FS, UK. [3]Department of Physics, Hanyang University, Seoul 04763, Republic of Korea. [4]Department of Physics, Chung-Ang University, 06974 Seoul, Republic of Korea. [5]Department of Physics, Ajou University, Suwon 16499, Republic of Korea.
*Corresponding author. Email: cg.oh.0404@gmail.com (C.-g.O.); sunwookim@hanyang.ac.kr (S.-W.K.); kunx@cau.ac.kr (K.W.K.); bm418@cam.ac.uk (B.M.); jwrhim@ajou.ac.kr (J.-W.R.)
†These authors contributed equally to this work.

## RESULTS AND DISCUSSION

### Quadratic band touching

A QBT refers to a scenario where two bands touch at a single point quadratically, as illustrated in Fig. 1A. In this work, we focus on the minimal situation in which the touching is well-isolated from other bands, allowing the low-energy physics to be accurately described by an effective two-band Hamiltonian. While multiband connectivity involving three or more bands can occur, such cases fall outside the scope of our current universality claim. In general, there are three types of QBT, as shown in Fig. 1A: (i) two bands with opposite-sign effective masses, (ii) two bands with the same-sign effective masses, and (iii) one of the touching bands has an infinite effective mass. This feature is frequently observed in condensed matter systems, such as bilayer graphene (*46–48*) and kagome materials (*49–52*). In kagome materials, one of the touching bands becomes extremely massive, as illustrated in the two rightmost panels of Fig. 1A. As most QBT semimetals exhibit isotropic band dispersion around the touching point, we focus on the isotropic QBT semimetals. The QBT point is geometrically characterized by the maximum value of the quantum distance, denoted by $d_{\mathrm{max}}$, between Bloch eigenvectors in the vicinity of the touching point, where the quantum distance between two quantum states $\psi_{\boldsymbol{k}_1}$ and $\psi_{\boldsymbol{k}_2}$ with momenta $\mathbf{k}_1$ and $\mathbf{k}_2$ is given by $d^2 = 1 - |\langle \psi_{\boldsymbol{k}_1} | \psi_{\boldsymbol{k}_2} \rangle|^2$ (*26*, *33–35*, *53*). A nonzero $d_{\mathrm{max}}$ arises when the Bloch wave function exhibits a discontinuity at the QBT point (*26*, *54*). This singularity can be visualized via the pseudospin texture, as shown in Fig. 1B, where the quantum distance between eigenvectors near the QBT point corresponds to the canting angle between the pseudospins representing these eigenvectors (*26*). Here, the pseudospin represents the two degrees of freedom, such as sublattices or real spins, inherent in the two-band model near the QBT point.

### Geometric properties of two-band systems

We first introduce several geometric concepts in an $N$-dimensional Hilbert space. The system is described by a complete set of quantum states $\{ |\psi_n(\Lambda)\rangle \}$ that depend smoothly on a set of real parameters $\Lambda = (\Lambda_1, \Lambda_2, \ldots)$, where $n \in \{1, \ldots, N\}$ is a quantum number, interpreted here as the band index. The Hilbert-Schmidt quantum distance for the $n$th band is defined as (*19*, *22*)

$$d^2_{\mathrm{HS},n}(\Lambda, \Lambda') = 1 - |\langle \psi_n(\Lambda) | \psi_n(\Lambda') \rangle|^2 \quad (1)$$

which is a dimensionless quantity ranging from 0 to 1 depending on the resemblance between the quantum states. If we consider the infinitesimal distance, we arrive at the quantum geometric tensor of the $n$th band

$$\mathcal{G}^n_{ij} = \left\langle \frac{\partial \psi_n}{\partial \Lambda_i} \middle| 1 - |\psi_n\rangle\langle\psi_n| \middle| \frac{\partial \psi_n}{\partial \Lambda_j} \right\rangle \quad (2)$$

$$= g^n_{ij}(\Lambda) - \frac{i}{2}\Omega^n_{ij}(\Lambda) \quad (3)$$

The symmetric part of the quantum geometric tensor $\mathrm{Re}\mathcal{G}^n_{ij} = g^n_{ij}(\Lambda)$ is the quantum metric tensor, and the antisymmetric part corresponds to the Berry curvature $\Omega^n_{ij}(\Lambda) = -2\mathrm{Im}\mathcal{G}^n_{ij}$ (*19*, *20*, *22*). The position vector in the parameter space $(\Lambda)$ usually takes the form of a crystal momentum $\mathbf{k}$ in condensed matter physics.

We now consider the geometric structures of a general two-band system, where the Hamiltonian is expressed as

$$H(\mathbf{k}) = h_0(\mathbf{k}) + \mathbf{h}(\mathbf{k}) \cdot \sigma \quad (4)$$

where $\sigma = (\sigma_x, \sigma_y, \sigma_z)$ denotes the Pauli matrices, while $h_0(\mathbf{k})$ and $\mathbf{h}(\mathbf{k})$ are smooth real-values functions. The pseudospin of the upper band is characterized as $s = \mathbf{h}/|\mathbf{h}|$. Using this pseudospin, the geometric quantities can be written as

$$d^2_{\mathrm{HS},n}(\mathbf{k}, \mathbf{k}') = \frac{1}{2}[1 - \mathbf{s}(\mathbf{k}) \cdot \mathbf{s}(\mathbf{k}')] \quad (5)$$

$$g^n_{ij}(\mathbf{k}) = \frac{1}{4}\partial_{k_i}\mathbf{s}(\mathbf{k}) \cdot \partial_{k_j}\mathbf{s}(\mathbf{k}) \quad (6)$$

$$\Omega^n_{ij}(\mathbf{k}) = -\frac{n}{2}\mathbf{s} \cdot \left(\partial_{k_i}\mathbf{s} \times \partial_{k_j}\mathbf{s}\right) \quad (7)$$

where $n$ is band index.

### Optical conductivity

On the basis of the Kubo formula, the interband conductivity for a 2D system is given by

$$\sigma_{ij}(\omega)=\frac{e^2}{\hbar}\int\frac{d^2k}{(2\pi)^2}\sum_{n,m}F_{nm}(\mathbf{k})\frac{i\epsilon_{mn}(\mathbf{k})A^i_{nm}(\mathbf{k})A^j_{mn}(\mathbf{k})}{\epsilon_{nm}(\mathbf{k})+\hbar\omega+i\eta} \quad (8)$$

where $F_{nm}(\mathbf{k})=f\left[\epsilon_n(\mathbf{k})\right]-f\left[\epsilon_m(\mathbf{k})\right]$, $f(\epsilon)=1/\left[1+e^{(\epsilon-\mu)/kT}\right]$ is the Fermi distribution function, $\epsilon_n(\mathbf{k})$ is the $n$th band energy, $\epsilon_{nm}(\mathbf{k})=\epsilon_n(\mathbf{k})-\epsilon_m(\mathbf{k})$, $\mu$ is the chemical potential, and $\eta$ is an infinitesimal real number resulting in a level broadening. The geometric factor $A^j_{nm}(\mathbf{k})$ is defined by

$$A^j_{nm}(\mathbf{k})=i\langle u_n(\mathbf{k})\,|\,\partial_{k_j}\,|\,u_m(\mathbf{k})\rangle \quad (9)$$

where $|u_{n\mathbf{k}}\rangle$ is the cell-periodic part of the Bloch wave function. We call it the Berry connection if $n=m$ and the interband Berry connection if $n\neq m$. Although $A^i_{nn}A^j_{nn}$ is gauge-dependent, $A^i_{nm}A^j_{mn}$ is gauge-invariant for $n\neq m$. For $n=m$, it does not contribute to the optical conductivity because $\epsilon_{nn}(\mathbf{k})=0$. Therefore, the optical conductivity is gauge-independent (see also Supplementary Text S1 for details). Note that this formulation assumes well-defined band quasiparticles so that the linear optical response is captured within the conventional band-theory (Fermi-liquid) framework; strongly correlated non–Fermi-liquid regimes are not addressed here.

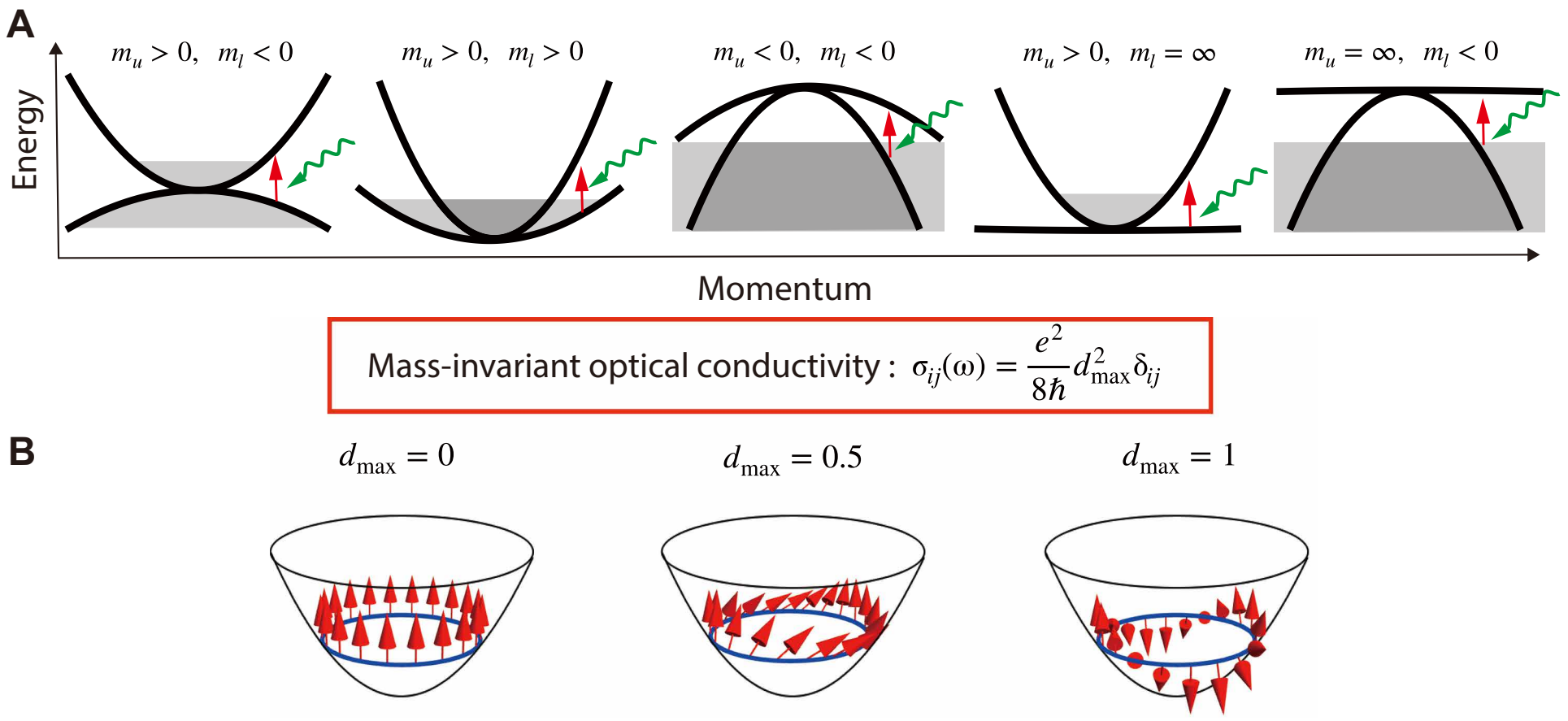

**Fig. 1. Schematics of mass invariant optical conductivity.** (**A**) A schematic illustration of the mass invariant optical conductivity $\sigma_{ij}$ for various systems with isotropic QBTs. In these systems, the optical conductivity is proportional to the square of the maximum quantum distance $d_{\mathrm{max}}$. The gray-shaded regions indicate the occupied states. (**B**) Pseudospin structures $[s_x(\mathbf{k}), s_y(\mathbf{k}), s_z(\mathbf{k})]$ of the isotropic QBT models for $d_{\mathrm{max}} = 0, 0.5,$ and 1.

For a two-band system with $\omega>0$, the real part of the optical conductivity is described as

$$\mathrm{Re}\left[\sigma_{ij}(\omega)\right]=\frac{\pi e^2}{\hbar}\int\frac{d^2\mathbf{k}}{(2\pi)^2}F_{lu}(\mathbf{k})\epsilon_{ul}(\mathbf{k})g^u_{ij}(\mathbf{k})\delta\left(\hbar\omega-\epsilon_{ul}\right) \quad (10)$$

where $u$ and $l$ denote upper and lower bands, respectively (see Supplementary Text S2 for the detailed derivation). This equation shows that the optical conductivity is determined by the band dispersions $\epsilon_n(\mathbf{k})$ and quantum metric tensor $g^u_{ij}(\mathbf{k})$.

## General isotropic QBT model

We consider a 2D continuum model describing two isotropic bands with a QBT point, where the lowest energy dispersions are given by quadratic band dispersions

$$\epsilon_{u/l}(\mathbf{k})=\frac{1}{2m_{u/l}}k^2 \quad (11)$$

where the effective mass $m_{u/l}$ can be either positive, negative, or infinite, as shown in Fig. 1A. Without loss of generality, we assume that the band touching occurs at $\mathbf{k}=0$. The most general Hamiltonian of such a model can be written with three parameters $(m_u, m_l, d_{\max})$ (*34*), as

$$\mathcal{H}_0(\mathbf{k})=\sum_\alpha h_\alpha(\mathbf{k})\sigma_\alpha \quad (12)$$

where $\sigma_\alpha$ represents the identity ($\alpha=0$) and Pauli matrices ($\alpha=x,y,z$). In addition, $h_\alpha(\mathbf{k})$ is a real quadratic function $h_0(\mathbf{k})=\left(1/M+2/m_l\right)\left(k_x^2+k_y^2\right)/4$, $h_x(\mathbf{k})=d_{\max}\left(1-d_{\max}^2\right)^{1/2}k_y^2/(2M)$, $h_y(\mathbf{k})=d_{\max}k_xk_y/(2M)$, and $h_z(\mathbf{k})=\left[k_x^2+\left(1-2d_{\max}^2\right)k_y^2\right]/(4M)$, where $1/M=1/m_u-1/m_l$. Here, $d_{\max}$ is the largest value of the quantum distance between all possible pairs of Bloch eigenvectors in the same band. Note that we consider the isotropic energy dispersion (i.e., $\epsilon_{u/l}\propto k^2$), which does not require the Hamiltonian itself to be rotationally symmetric. While the geometric quantity $d_{\max}$ represents the strength of the interband coupling, it does not manifest in the band dispersion as the band dispersion is entirely determined by only two parameters $(m_u, m_l)$. QBT systems provide the minimal continuum model– $\epsilon\propto k^N$ with $N=2$ in which $d_{\max}$ is a freely tunable, purely geometric parameter (ranging from 0 to 1). This unique tunability lets us isolate and probe the effects of quantum geometry, something impossible in linear ($N=1$) models where $d_{\max}$ is fixed (*55*).

As $d_{\max}$ increases, the pseudospin canting becomes more pronounced, resulting in a larger maximum relative angle between pseudospins. Ultimately, a complete winding structure emerges in the pseudospin texture when $d_{\max}=1$, as shown in Fig. 1B. Since the geometric properties of the two-band model can be translated into the pseudospin structures as derived from Eqs. 5 to 7, $d_{\max}$ can be considered the parameter that governs the geometry of the system. The quantum metric of the system is given by

$$g^n_{xx}(\mathbf{k})=d_{\max}^2\frac{k_y^2}{k^4},\ g^n_{yy}(\mathbf{k})=d_{\max}^2\frac{k_x^2}{k^4}$$
$$g^n_{xy}(\mathbf{k})=g^n_{yx}(\mathbf{k})=-d_{\max}^2\frac{k_xk_y}{k^4} \quad (13)$$

The Berry curvature is zero in this QBT system as studied previously (*55*). Notably, the band masses do not contribute to the quantum geometric tensor, which is a critical factor underlying the universal optical conductivity. In contrast, for massive Dirac fermion systems, the quantum metric is directly affected by the mass, resulting in an optical conductivity that is strongly dependent on the band structure (*41*, *42*). A detailed comparison of the quantum metrics for massive and massless Dirac fermions is provided in table S1 of Supplementary Text S8. When $d_{\max}=1$, other geometric quantities, such as the winding number (*33*) and the Euler invariant, applicable in the case $m_u=-m_l$ with $C_2T$ symmetry (*56*, *57*), can also be used.

We calculate the optical conductivity of this model for $\omega>0$ at zero temperature. The real part of the conductivity is given by

$$\mathrm{Re}\left[\sigma_{ij}(\omega)\right]=\frac{e^2}{\hbar}\frac{d_{\max}^2}{8}\delta_{ij}\Theta(\hbar\omega-\mu^*) \quad (14)$$

where $\Theta(x)$ is the Heaviside step function and $\mu^*=m_u\mu/M$ with a chemical potential $\mu$ (See Supplementary S4 for the derivation). When $\omega<\mu^*$, electrons in the lower band cannot transition to the upper band because the final states are already occupied. The conductivity is flat for $\omega>\mu^*$, and its intensity depends solely on the maximum quantum distance $d_{\max}$, independent of the band structure. The $d_{\max}^2$ factor implies that finite optical conductivity occurs only when there is an interband coupling between touching bands. Namely, even if two irrelevant bands approach and eventually touch accidentally, the optical conductivity remains zero if $d_{\max}$ is zero.

The origin of this mass-invariant and frequency-independent universal conductivity lies in a remarkable cancellation that occurs within the optical conductivity integral (Eq. 10), a feature unique to 2D isotropic QBT systems. The integrand consists of three key parts: the energy dispersion ($\epsilon_{ul}\propto k^2/M$), the quantum metric ($g_{ij}\propto 1/k^2$), and the 2D k-space measure ($d^2k\propto kdk$). During the integration, a cascade of cancellations occurs. The $k^2$ dependence from the energy dispersion is precisely negated by the $1/k^2$ scaling of the quantum metric. Subsequently, the mass-dependent prefactor $(1/M)$ from the energy term is canceled by the mass dependence arising from the integration over the delta function, which involves the constant density of states (proportional to $M$). This process simultaneously eliminates all dependence on both the effective masses and the transition frequency (which only serves to a threshold), leaving a constant conductivity determined solely by the geometric parameter $d_{\max}$.

For different dispersions (e.g., Supplementary Text S10) or in other dimensions (e.g., Supplementary Text S12), these perfect cancellations do not occur, making this phenomenon specific to 2D isotropic QBT systems.

### Lattice model

To validate the applicability of our continuum model to lattice systems, we consider the following lattice Hamiltonian

$$\mathcal{H}_{lat}(\mathbf{k}) = \sum_{\alpha} g_{\alpha}(\mathbf{k})\sigma_{\alpha} \tag{15}$$

where $g_0(\mathbf{k}) = 0$, $g_x(\mathbf{k}) = 2d_{\max}\sqrt{1-d_{\max}^2}\left(1-\cos k_y\right)$, $4d_{\max}\sin$ $4d_{\max}\sin\left(k_x/2\right)\sin\left(k_y/2\right)$, and $2-2d_{\max}^2-\cos k_x+\left(2d_{\max}^2-1\right)\cos k_y$. The enegy eigenvalues of this model are given as $E_{\pm} = \pm\left(2-\cos k_x-\cos k_y\right)$, as shown in Fig. 2A. A key feature of this theoretical lattice model is that the parameter $d_{\max}$ is constructed to be independent of the terms that determine the band dispersion. This conceptual separation, while not implying that $d_{\max}$ is a tunable knob in real materials, allows the model to clearly isolate and illustrate the effects of quantum geometry on the optical conductivity. Performing a $\mathbf{k}\cdot\mathbf{p}$ expansion of this model at the $\Gamma$ point up to quadratic order yields the isotropic QBT model described in Eq. 12 with $m_u = -m_l = 1$.

Figure 1B shows the optical conductivity for $d_{\max} = 0, 0.5$, and 1. The solid lines represent the results from the continuum model, while the markers, obtained by calculating the Kubo formula, correspond to the lattice model. The agreement between the continuum model and lattice results at low energies confirms the validity of our theoretical framework.

As another example, we consider the kagome lattice model, including only nearest-neighbor hopping. The corresponding Hamiltonian is given in Supplementary Text S5. When analyzing the lowest two bands (Fig. 2C), a $\mathbf{k}\cdot\mathbf{p}$ expansion near the band-touching point reveals that the system corresponds to an isotropic QBT model with $d_{\max} = 1$ and an infinite effective mass $m_l = \infty$ (*26*, *35*). Calculating the optical conductivity for this system yields a value of $e^2/(8\hbar)$, as shown in Fig. 3D, which is consistent with the prediction for $d_{\max} = 1$. This result further supports the validity of our theoretical framework, demonstrating its applicability to systems with both finite and infinite effective masses.

Examining Fig. 2 (B and D), we observe that as the frequency $\omega$ increases, deviations between the lattice and continuum models emerge. Specifically, the lattice model results no longer exhibit flat optical conductivity. These deviations arise from higher-order corrections in the lattice model that are not accounted for in the continuum approximation. The effects of the higher-order corrections to the optical conductivity are examined in Supplementary Text S6.

### Symmetry constraints on optical conductivity

In the preceding analysis, we only assumed an isotropic energy dispersion (i.e., $\epsilon_{u/l} \propto k^2$), which is a condition on the energy eigenvalues alone. We now impose additional symmetries on the Hamiltonian itself.

For a 2D spinless QBT system that preserves both $n$-fold rotational symmetry $C_n$ and time-reversal symmetry $\mathcal{T}$, the maximum quantum distance $d_{\max}$ around the band-touching point is quantized to either 0 or 1. Detailed derivations are provided in Supplementary Text S9. Consequently, the optical conductivity calculated using Eq. 14 is quantized to either 0 or $e^2/(8\hbar)$ when the touching occurs at high-symmetry points with $C_n$ rotational symmetry. This remarkable quantization of optical conductivity, constrained by symmetry, is independent of the band masses at the touching point.

Since the case $d_{\max} = 0$ corresponds to the absence of interaction between the upper and lower bands, resulting in zero optical conductivity, we neglect this scenario as such accidental band touching are unlikely in real materials. We demonstrate below that a wide range of QBT materials with $C_n$ and $\mathcal{T}$ symmetries exhibit nonzero optical conductivity, corresponding to the case where $d_{\max} = 1$.

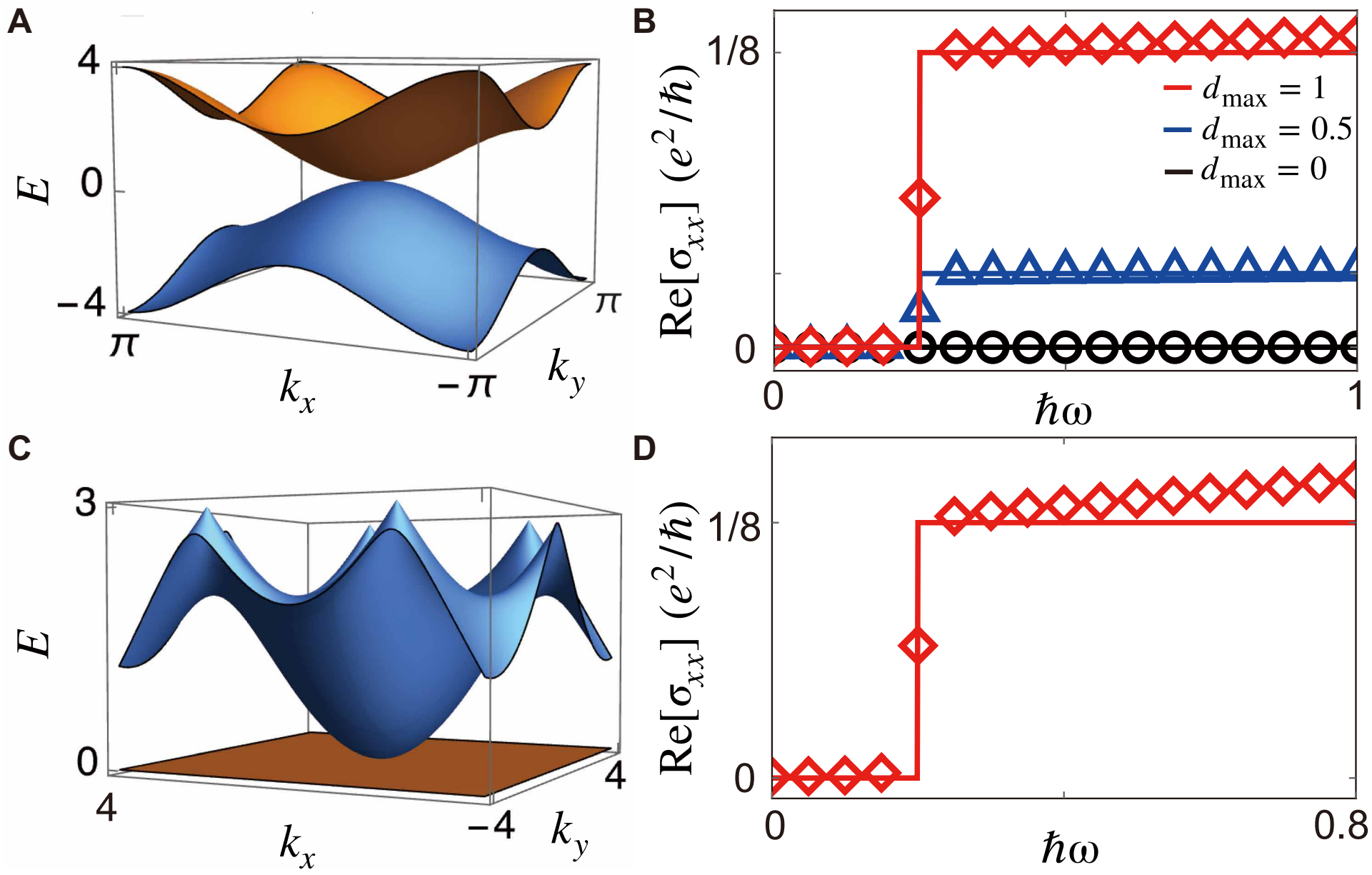

**Fig. 2. Mass invariant optical conductivity in lattice models.** (**A**) Band dispersions of the lattice model $\mathcal{H}_{lat}$ as given in Eq. 15. (**B**) Frequency $\omega$ dependence of the real part of the optical conductivity $\mathrm{Re}\left[\sigma_{xx}\right]$ for $d_{\max} = 0$ (black), 0.5 (blue), and 1 (red). (**C**) The lowest two band dispersions of the kagome lattice model (see Supplementary Text S5 for details). (**D**) Frequency dependence of the real part of the optical conductivity. The calculations in (B) and (D) are performed at $\mu^* = 0.2$. The solid lines represent the results from the isotropic QBT model in Eq. 14, while the markers denote those from the lattice models.

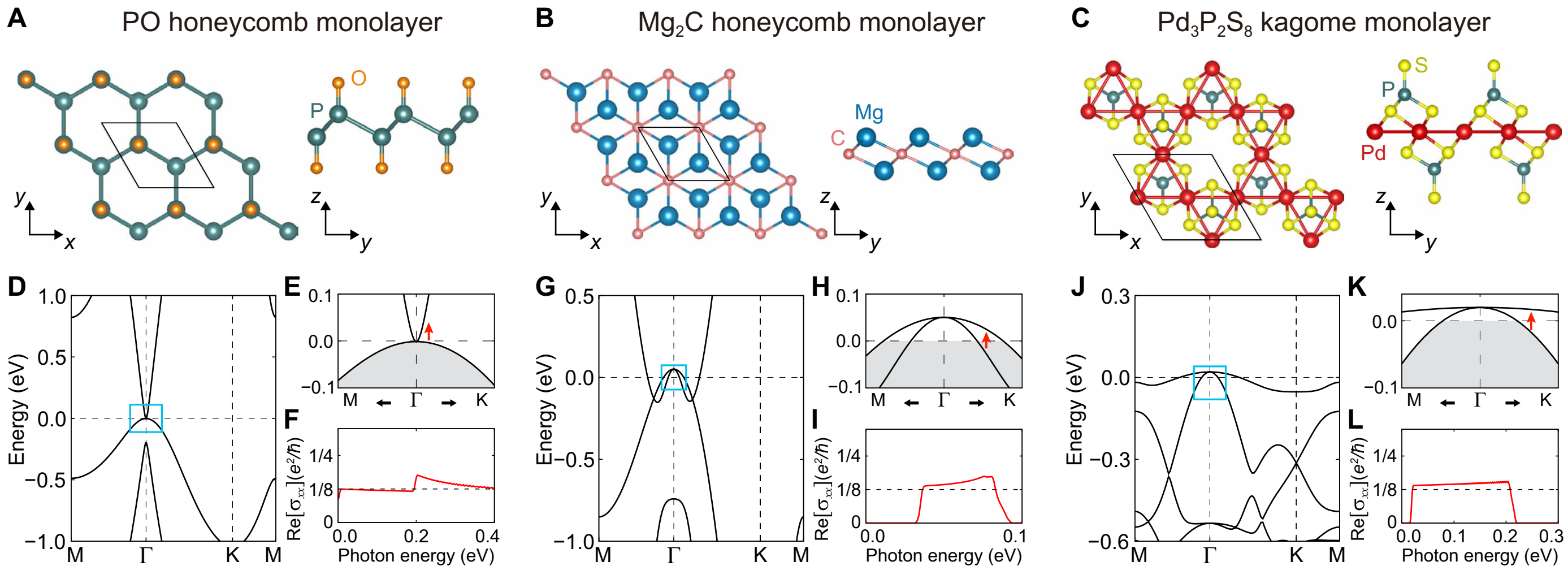

**Fig. 3. Universal optical conductivity in various 2D QBT materials.** (**A** to **C**) Atomic structures of (A) PO, (B) $Mg_2C$, and (C) $Pd_3P_2S_8$ monolayers, showing the top view (left) and the side view (right). (**D** to **F**) PO monolayer with 2% tensile strain: (D and E) DFT band structures and (F) the real part of the interband optical conductivity $\mathrm{Re}[\sigma_{xx}]$. (**G** to **I**) Slightly electron-doped $Mg_2C$ monolayer: (G and H) DFT band structures and (I) the real part of the interband optical conductivity $\mathrm{Re}[\sigma_{xx}]$. (**J** to **L**) Slightly hole-doped $Pd_3P_2S_8$ monolayer: (J and K) DFT band structures and (L) the real part of the interband optical conductivity $\mathrm{Re}[\sigma_{xx}]$.

## Manifestation of universal optical conductivity in various materials

To confirm the model predictions, we perform first principles density functional theory (DFT) calculations for various 2D QBT materials. We find that these materials exhibit the quantized optical conductivity value of $\sigma = e^2/(8\hbar)$ near QBT points, irrespective of their band masses. We illustrate this with several nonmagnetic materials that exhibit $C_3$ rotational symmetry but differ in their band masses near the touching point: honeycomb PO (*58*), honeycomb $Mg_2C$ (*59*), and kagome $Pd_3P_2S_8$ (*60*) monolayers (Fig. 3).

The PO monolayer hosts a QBT at the Γ point with opposite signs of band masses, i.e., $m_u > 0$ and $m_l < 0$, under 2.0% tensile strain (Fig. 3, D and E). The calculated optical conductivity shows the quantized value of $\sigma = e^2/(8\hbar)$ up to 0.2 eV (Fig. 3F). It increases from the quantized value around 0.2 eV because of additional interband transitions originating from the occupied band below the quadratic touching bands.

The $Mg_2C$ monolayer is characterized by a QBT with the same sign of band masses, where $m_u < 0$ and $m_l < 0$ (Fig. 3, G and H). To examine the interband optical conductivity originating exclusively from interband transitions between the two bands near the touching at the Γ point, we shift the Fermi level upward by 0.1 eV, corresponding to electron doping. The calculated optical conductivity displays the quantized value of $\sigma = e^2/(8\hbar)$ (Fig. 3I). It exhibits a slightly increasing behavior as photon energy increases due to high-order effects (see Supplementary Text S6).

The kagome $Pd_3P_2S_8$ monolayer represents an extreme case of a QBT, where the effective mass of one of the bands is extraordinarily large near the touching point at Γ (Fig. 3, J and K). This large band mass of the upper touching band is attributed to the flat band of the Pd kagome lattice, which spans the entire Brillouin zone. Note that the flat band is dispersive because of long-range hoppings beyond the nearest-neighbor hopping. The calculated interband optical conductivity displays the quantized value of $\sigma = e^2/(8\hbar)$ (Fig. 3L), consistent with the prediction from the ideal kagome lattice model in Fig. 2D. Since this kagome monolayer is insulating in its pristine state, we introduce moderate hole doping to explore interband transitions between the two touching bands at Γ when calculating the optical conductivity.

The universal optical conductivity of QBT materials remains remarkably robust against small bandgap openings when $d_{\max} = 1$. This robustness is reflected in the optical conductivity formula derived for a slightly gapped QBT (see Supplementary Text S7 for details). We illustrate this behavior in the strained honeycomb Bi monolayer (Fig. 4), which hosts a QBT at Γ with opposite signs of band masses under 2.1% tensile strain. This QBT again gives rise to the universal optical conductivity of $\sigma = e^2/(8\hbar)$. When the strain is reduced, the band touching is lifted, leading to an insulating state with an indirect gap of 27 and 113 meV for 1.9 and 1.0% strain, respectively. Notably, the optical conductivity remains largely flat near the value of $\sigma = e^2/(8\hbar)$, with a small peak developing at low photon energies as the gap size increases. This robustness highlights the mass-invariant nature of the universal optical conductivity in QBT materials.

To demonstrate the robustness of our theory against realistic anisotropies, we revisit the optical conductivity of AB-stacked bilayer graphene. It has been shown to exhibit the universal optical conductivity of $\sigma = e^2/(2\hbar)$ (*61*, *62*), corresponding to $\sigma = e^2/(8\hbar)$ per spin and per valley. We attribute this to the quantum geometry value $d_{\max} = 1$, which is enforced by the symmetry of the system. While it is a quintessential QBT semimetal, it exhibits a small trigonal warping (*48*) that splits the touching point into four linear cones with an energy scale of ~3 meV (Fig. 5). Despite this splitting, our calculations confirm that the universal optical conductivity remains remarkably robust, showcasing that our findings are not confined to idealized isotropic models. The only exception is a small additional peak at this energy scale, which can be suppressed either at a photon frequency beyond 3 meV or by slightly shifting the chemical potential (±2 meV). This further underscores the robustness of the flat universal optical conductivity, as has already been experimentally observed in AB-stacked bilayer graphene (*62*).

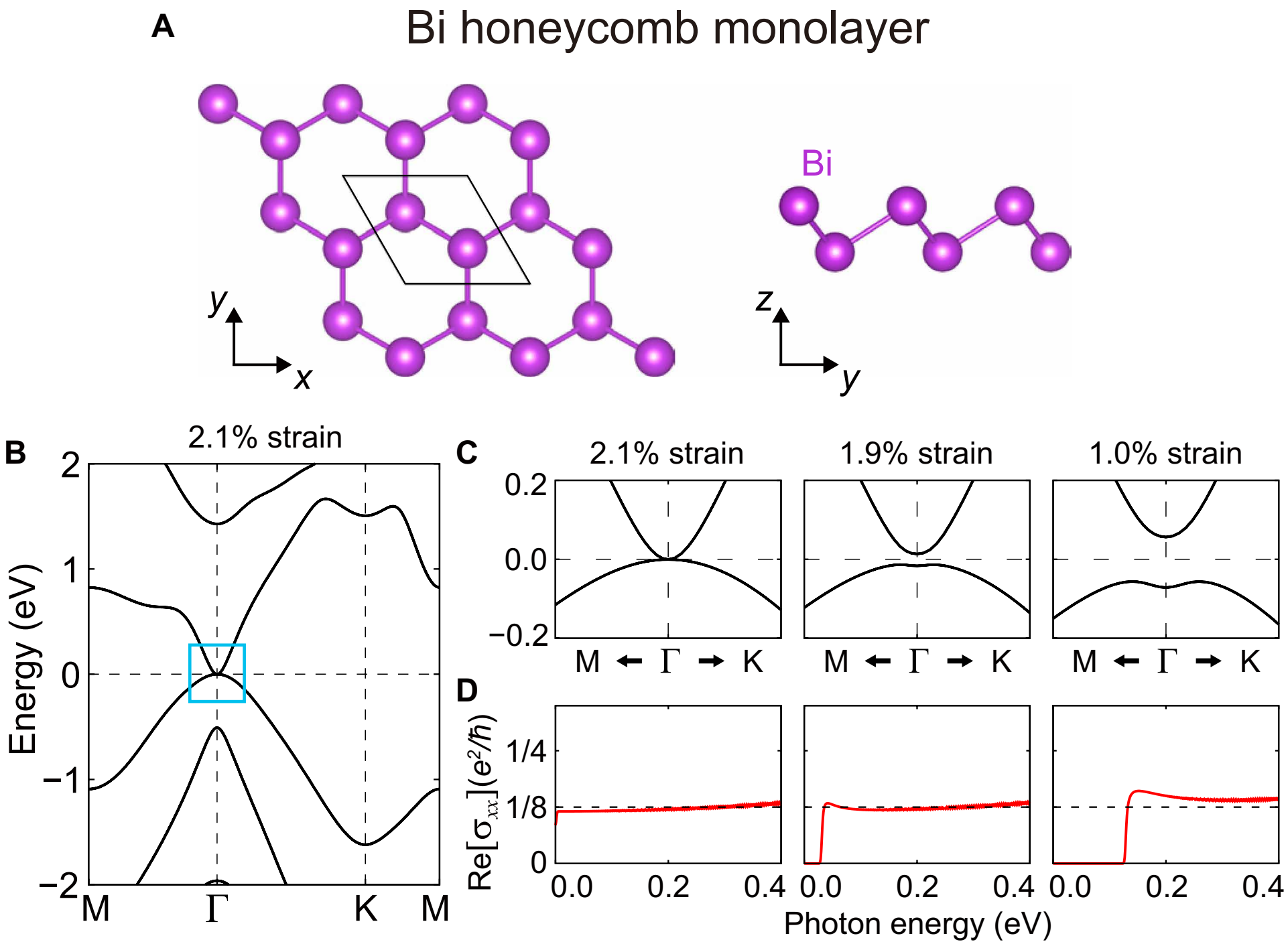

**Fig. 4. Robustness of universal optical conductivity against small gap opening.** (**A**) Atomic structures of Bi monolayer, showing the top view (left) and the side view (right). (**B** and **C**) DFT band structures as a function of biaxial tensile strain. (**D**) Corresponding the real part of the interband optical conductivity $\mathrm{Re}[\sigma_{xx}]$ as a function of the strain.

The universal optical conductivity in bilayer graphene, previously studied in the context of N-layer graphene systems, is often attributed to the system's chiral symmetry. Our work, however, provides a complementary and more general perspective based on quantum geometry. From our framework, bilayer graphene is a prime example of a 2D QBT system where the universal conductivity arises from a quantized geometric quantity, $d_{\mathrm{max}} = 1$. While bilayer graphene has chiral symmetry, our analysis reveals that the more fundamental requirement for this universal conductivity in QBT systems is the quantum geometry enforced by $C_3$ and $\mathcal{T}$ symmetries. This mechanism is more general as it remains valid even for systems that lack the chiral symmetry. The strength of our framework is its ability to explain why the same universal conductivity appears in other QBT materials (such as $Mg_2C$ and $Pd_3P_2S_8$) that have the necessary $C_3$ and $\mathcal{T}$ symmetries but do not share the specific chiral symmetry of bilayer graphene.

## Experimental feasibility

We finally discuss the experimental feasibility of observing the universal optical conductivity in a broader class of materials. Our predictions apply to generic $C_n$ and $\mathcal{T}$ symmetric 2D materials with an isotropic QBT. Among the various stable material examples discussed, in addition to the well-studied bilayer graphene, $Pd_3P_2S_8$ and Bi monolayers have been experimentally realized (*60*, *63*), offering promising opportunities to test our predictions. We expect that structures featuring a QBT at the Fermi level without any other states nearby, like the strained Bi and PO honeycomb monolayers, would be ideal for observing the flat universal optical conductivity. This guarantees the suppression of the intraband Drude response (see fig. S2 in Supplementary Text S11), which could otherwise hinder the observation of the flat optical conductivity at low photon energies. Moreover, as illustrated above, QBT materials with a possible small gap opening are also allowed to exhibit universal optical conductivity. This robustness, combined with the first-order nature of linear optical conductivity—which can be directly measured using a simple setup with high-intensity outputs—facilitates accessible pathways for experimental observation. We note that although higher-order terms beyond the quadratic order at a band-touching point generally suppress the flatness of the universal conductivity, the value of $\sigma = e^2/(8\hbar)$ at the onset of an interband transition is retained (see Supplementary Text S6). Overall, our prediction for the universal optical conductivity in the linear response regime offers a compelling way to directly probe quantum geometry.

We demonstrate that the optical conductivity in 2D QBT systems is universally determined by the quantum geometry, characterized by the maximum quantum distance at the band-touching point, and is independent of the details of the band structure. Specifically, the optical conductivity is given by $\sigma = \left(e^2/8\hbar\right)d_{\mathrm{max}}^2$, consisting only of the fundamental constants $e$ and $\hbar$ and of the quantum geometric quantity $d_{\mathrm{max}}$. Moreover, we show that the optical conductivity is quantized to $\sigma = e^2/(8\hbar)$ when the system respects time-reversal and rotational symmetries. These results hold across analytical models, tight-binding lattice models, and first-principles calculations of several 2D materials. While our analytical framework is based on a two-band model, its conclusions are expected to hold for realistic multiband systems where the QBT is well-isolated from other bands. In such cases, the low-energy physics is accurately captured by an effective two-band model. The strong agreement between our theory and the first-principles DFT results for various materials—which inherently include all bands—confirms that the low-energy physics is dominated by the two touching bands in these systems.

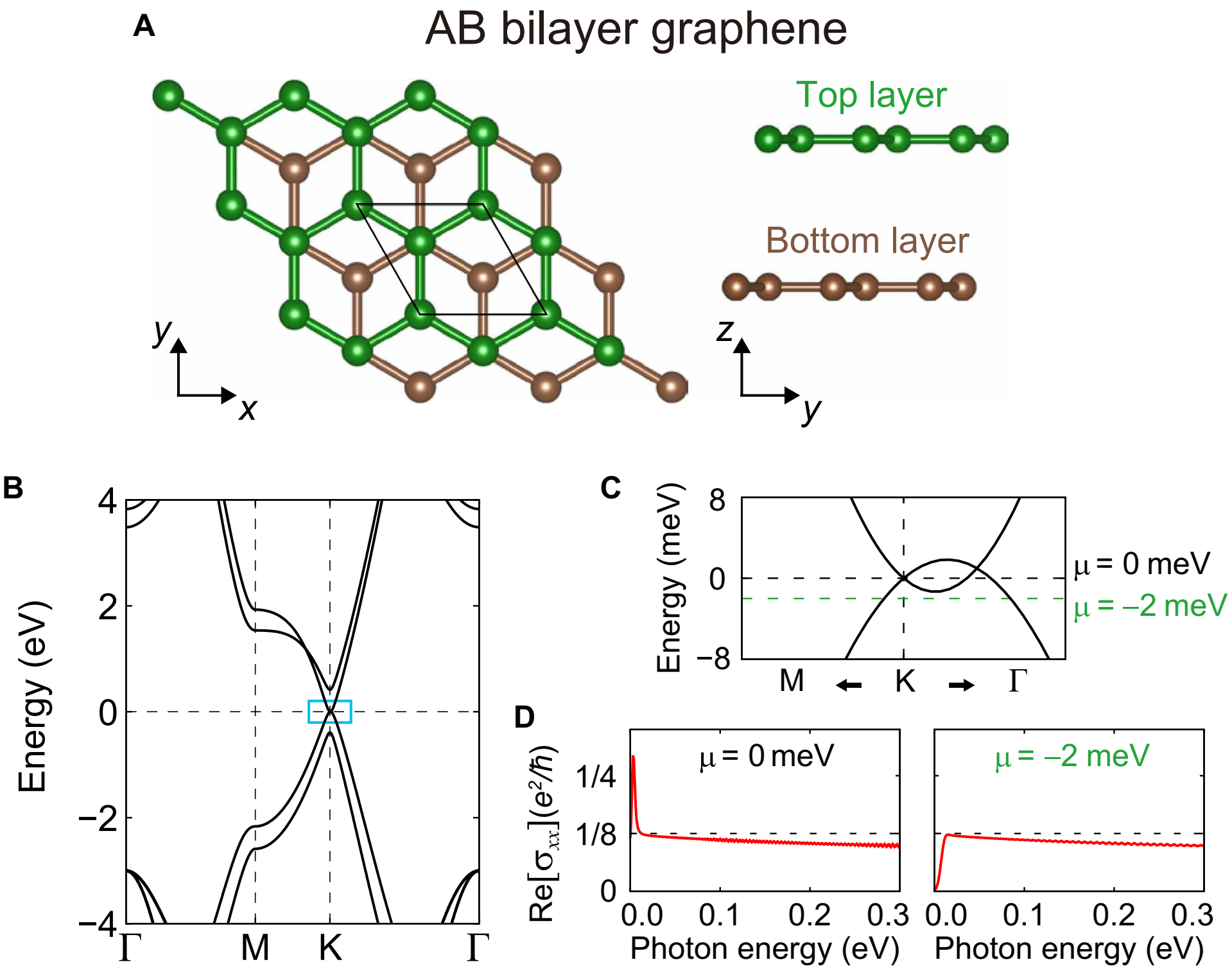

**Fig. 5. Robustness of universal optical conductivity against an anisotropy.** (**A**) Atomic structures of AB-stacked bilayer graphene, showing the top view (left) and the side view (right). (**B** and **C**) DFT band structures. (**D**) The real part of the interband optical conductivity $\mathrm{Re}[\sigma_{xx}]$ at two different chemical potential $\mu = 0$ and $-2$ meV.

However, we recognize that in systems where the QBT is not well-isolated or involves threefold or higher degeneracies, the minimal two-band framework may break down. In addition, our current formulation based on the Kubo formula assumes the existence of well-defined band quasiparticles within the conventional Fermi-liquid regime. Extending the geometric formulation to such genuinely multiband degeneracies and beyond-quasiparticle regimes, and identifying whether analogous geometric universalities can emerge there, constitutes an interesting direction for future work.

Our finding of a universal optical conductivity, solely determined by quantum geometric properties, adds to the rare examples of universal behaviors in condensed matter physics such as the quantum Hall conductance in 2D systems, the minimal conductivity (*64*), and the universal optical conductivity (*61*, *65*, *66*) of graphene. This universality is of a purely geometric origin, in contrast to topological phenomena like the quantum Hall effect. The QBT systems we study are topologically trivial with zero Berry curvature, yet they exhibit a nontrivial quantum geometry characterized by $d_{\mathrm{max}}$. This demonstrates that the quantum metric, distinct from the Berry curvature, can independently govern a quantized physical observable, highlighting a previously unexplored route to universal phenomena in condensed matter systems.

This universal optical conductivity, because of its linear nature, enables direct extraction of geometric properties of Bloch wave functions using standard optical measurements, in contrast to the complex and challenging spectroscopic techniques required for higher-order optical conductivity measurements. Moreover, our proposal provides further experimental advantages. Conventional approaches to probing quantum geometry often involve integrated optical quantities (*31*, *41*), such as $\int_0^\infty d\omega\sigma(\omega)/\omega$, which require broadband frequency measurements that are technically demanding. These integrated quantities also obscure the local momentum-resolved structure of quantum geometry, yielding only global information over the Brillouin zone. In contrast, QBT systems offer a rare exception: Knowledge of a single quantity $d_{\mathrm{max}}$, accessible through relatively low-frequency measurements, is sufficient to reconstruct the full local quantum geometric tensor at the band-touching point, as shown in Eq. 13. This accessibility makes our approach not only conceptually powerful but also experimentally feasible.

We have focused on the QBT model with $C_n$ and $T$ symmetries under isotropic conditions. However, when $n = 2$ or 4, these symmetries can permit anisotropic quadratic terms. One example of such an anisotropic QBT is described by the Hamiltonian $H_{\mathrm{aniso}} = a_0 k^2 \sigma_0 + a\left(k_x^2 - k_y^2\right)\sigma_x + 2bk_xk_y\sigma_y$. This system retains $d_{\mathrm{max}} = 1$, as detailed in Supplementary Text S10. If $a = b$, it corresponds to the isotropic QBT Hamiltonian described in Eq. 12 with $d_{\mathrm{max}} = 1$. A straightforward calculation yields the optical conductivity given by $\mathrm{Re}\left[\sigma_{xx}(\omega)\right] = e^2\left(a^2 + b^2\right)/(16\hbar ab)$, which depends on the shape of the bands. When $a = b$, the optical conductivity simplifies to $e^2/(8\hbar)$, consistent with the universal value for isotropic QBT systems. Other types of anisotropic QBT cases are discussed in the Supplementary Materials.

While our discussion primarily focuses on the 2D case, quantum geometry also significantly influences optical conductivity in 3D isotropic QBT systems. Unlike in the 2D case, the intensity of optical conductivity in 3D systems depends on the band structure, such as effective masses. However, similar to the 2D case, the contribution from quantum geometry is still encoded in the optical conductivity (see Supplementary Text S12). Thus, even in 3D, measuring optical conductivity can serve as an effective way to probe the geometric properties of the system, providing valuable insights into quantum geometry without requiring more complex spectroscopic techniques.

## METHODS

### Electronic structure calculations

We perform DFT calculations using the Vienna ab initio simulation package (VASP) (*67*, *68*) implementing the projector-augmented wave method (*69*). We approximate the exchange correlation functional with the generalized-gradient approximation of Perdew-Burke-Ernzerhof (*70*). We use a kinetic energy cutoff for the plane wave basis of 600 eV and a Gaussian smearing of 0.02 eV. We use Γ-centered **k**-point grids with a **k**-spacing of less than 0.1 $\text{Å}^{-1}$. All the structures are optimized until the forces are below 0.001 eV/Å. We optimize the lattice constants of the monolayer structures, except when the experimental lattice parameters are known, as in the case of $Pd_3P_2S_8$ (*71*) and Bi (*72*). Each monolayer structure is simulated using a periodic supercell with a vacuum spacing of 20 Å in the direction perpendicular to the plane. Spin-orbit coupling is included for the Bi monolayer. The interband optical conductivity is calculated using the Kubo-Greenwood formula, as implemented in the WANNIER90 package (*73*) with dense **k**-point grids with a spacing of less than 0.002 $\text{Å}^{-1}$.

## Supplementary Materials

**This PDF file includes:**
Supplementary Text S1 to S12
Figs. S1 and S2
Table S1
References

**Acknowledgments**

**Funding:** We thank P. Törmä for useful discussions. C.-g.O. was supported by Q-STEP, WINGS Program, the University of Tokyo, and JSPS KAKENHI grant number JP25KF0186. S.-W.K. was supported by a Leverhulme Trust Early Career Fellowship (ECF-2024-052), by a UKRI Future Leaders Fellowship (MR/V023926/1) and by the research fund of Hanyang University (HY-202600000001189). K.W.K. acknowledges financial support from the Basic Science Research Program through the National Research Foundation of Korea (NRF) funded by the Ministry of Education (no. RS-2025-00521598) and the Korean Government (MSIT) (no. 2020R1A5A1016518). B.M. was supported by a UKRI Future Leaders Fellowship (MR/V023926/1) and by the Gianna Angelopoulos Programme for Science, Technology, and Innovation. J.W.R. was supported by the National Research Foundation of Korea (NRF) Grant funded by the Korean government (MSIT) (grant nos. 2021R1A2C1010572 and RS-2023-NR068116) and the Ministry of Education (grant no. RS-2023-00285390). The computational resources were provided by the Cambridge Tier-2 system operated by the University of Cambridge Research Computing Service and funded by EPSRC (EP/P020259/1), and by the UK National Supercomputing Service ARCHER2, for which access was obtained via the UKCP consortium and funded by EPSRC (EP/X035891/1). **Author contributions:** C.-g.O., S.-W.K., and J.-W.R. conceived the project. C.-g.O. and S.-W.K. performed the theoretical analysis and first-principles calculations. K.W.K. and B.M. contributed to the interpretation of the results. All authors discussed the results and contributed to the writing of the manuscript. **Competing interests:** The authors declare that they have no competing interests. **Data, code, and materials availability:** All data and code needed to evaluate and reproduce the results in the paper are present in the paper and/or the Supplementary Materials. The VASP code used in this study is a commercial electronic structure modeling software available from https://vasp.at. This study did not generate new materials.

Submitted 14 April 2025
Accepted 24 April 2026
Published 5 June 2026
10.1126/sciadv.ady2033

# Mass-invariant universal optical conductivity from quantum geometry

Chang-geun Oh, Sun-Woo Kim, Kun Woo Kim, Bartomeu Monserrat, and Jun-Won Rhim